\documentclass[aps,prd,eqsecnum,preprint,amsmath,showpacs]{revtex4}

\usepackage{color,graphicx}
\usepackage{amsfonts}
\usepackage{amssymb}

\begin{document}


\title{Uniqueness of static spherically symmetric vacuum solutions
in the IR limit of Ho\v{r}ava-Lifshitz gravity}

\author{Tomohiro Harada\footnote{Electronic
address:harada@scphys.kyoto-u.ac.jp}, Umpei Miyamoto
and Naoki Tsukamoto}
\affiliation{
Department of Physics, Rikkyo University, Tokyo 171-8501, Japan
}
\date{\today}

\begin{abstract}                
We investigate static spherically symmetric vacuum solutions 
in the IR limit of projectable nonrelativistic quantum gravity,
including the renormalisable quantum gravity
recently proposed by Ho\v{r}ava.  
It is found that the projectability condition plays an important role. 
Without the cosmological constant, the spacetime is 
uniquely given by the Schwarzschild solution.
With the cosmological constant, 
the spacetime is uniquely given by the Kottler 
(Schwarzschild-(anti) de Sitter) solution for 
the entirely vacuum spacetime. However, in addition to the Kottler solution,
the static spherical and hyperbolic universes are uniquely
admissible for the locally empty region, for the positive and negative
cosmological constants, respectively, if its nonvanishing contribution 
to the global Hamiltonian constraint can be compensated
by that from the nonempty or nonstatic region. 
This implies that static spherically symmetric 
entirely vacuum solutions would not admit the 
freedom to reproduce the observed flat 
rotation curves of galaxies. On the other hand, 
the result for locally empty regions implies that 
the IR limit of nonrelativistic quantum gravity
theories does not simply recover general relativity
but includes it.
\end{abstract}


\maketitle

\section{I. ~INTRODUCTION}

Recently, Ho\v{r}ava~\cite{horava2009_lifshitz} proposed a deeply
nonrelativistic quantum gravity theory, which is at a Lifshitz point with 
dynamical critical exponent $z=3$ in 3+1 dimensions
in the UV limit and apparently reduced to general relativity
in the IR limit. This theory is called the Ho\v{r}ava-Lifshitz gravity.
This is power-counting renormalisable and intended to be ghost-free.
The phenomenological aspects of this theory and its variants have been 
intensively studied, including black holes~\cite{lmp2009,park2009,kehagias_sfetsos2009,kiritsis_kofinas2009_blackhole}, 
cosmological
implications~\cite{kiritsis_kofinas2009,mukohyama2009,takahashi_soda2009}
and gravitational waves~\cite{svw2009_letter,svw2009_full}
(see e.g.~\cite{svw2009_full} and references therein).

Among the ingredients of Ho\v{r}ava-Lifshitz gravity are the 
detailed balance condition and the projectability condition. 
The detailed balance condition strongly restricts the form of 
the action. Although Ho\v{r}ava originally assumes this condition
from a renormalisability point of view, several authors 
subsequently abandon
this condition and extend the theory to more general class of 
actions~(see e.g.~\cite{svw2009_letter,svw2009_full}). 

The projectability condition is very intriguing 
because it seems to characterise 
nonrelativistic theories. Under this condition, the Hamiltonian 
constraint is nonlocal and obtained by the spatial integral.
As a result, the Hamiltonian and momentum constraints constitute
a closed algebra. This feature appears to be suitable for quantisation
compared to the notorious non-closed algebra for general 
relativity. Although this condition was often neglected in the early
stage of phenomenological studies, 
it was recently shown~\cite{mukohyama2009} that 
this condition in the IR limit results in the emergence
of pressureless fluid additional to general relativity and 
argued that this can play a role of cold dark matter.
On the other hand, it has been generally accepted that the rotation curves 
of galaxies are flatter than expected from the 
luminous matter distribution~\cite{deblock_bosma_2002}.
This observational fact suggests that the so-called 
galactic dark matter is responsible for the considerable 
fraction of the total mass of galaxies. 

In this article, we study static spherically symmetric vacuum
solutions of the IR limit of nonrelativistic quantum gravity
theories, taking properly into account the projectability condition.
We show that as in general relativity such solutions are 
uniquely given by the Kottler (Schwarzschild-(anti) de Sitter)
solutions. This implies that 
vacuum solutions would not be capable of maintaining the flat 
rotation curves of galaxies. On the other hand, we also show  
that static solutions can be locally described by three-sphere or
three-hyperboloid if the spacetime is not entirely but only locally 
empty. This indicates that nonrelativistic quantum gravity theories
would not simply recover general relativity in their IR limit.
We retain both the gravitational constant $G$ and the speed of light $c$.
\section{II. ~FIELD EQUATIONS}

The field variables of the Ho\v{r}ava-Lifshitz gravity
and other nonrelativistic quantum gravity theories
are the spatial metric $g_{ij}$ as well as
the lapse $N$ and the shift $N^{i}$.
The line element in the 
four dimensional spacetime manifold is given in terms 
of these geometrical quantities as
\begin{equation}
ds^{2}=-N^{2}c^{2}dt^{2}+g_{ij} (dx^{i}+N^{i}dt)
(dx^{j}+N^{j}dt).
\end{equation}
The action of nonrelativistic quantum gravity is constructed so that 
it is covariant under the foliation-preserving diffeomorphisms
$\delta t=f(t), \delta x^{i}=\zeta ^{i}(t,x^{j})$.
That is, the theory has no general covariance but the part of it.
We focus on the IR limit of the theory, where we can neglect the 
higher derivative terms in the action. 
We will use the curvature tensors for the spatial metric $g_{ij}$.
We consider 
the following action~\cite{horava2009_lifshitz,kiritsis_kofinas2009,mukohyama2009}:
\begin{equation} S_{g}=\int
dtd^3x\sqrt{g}N \left[\alpha (K_{ij}K^{ij}\!-\lambda K^2) + \xi
R + \sigma \right] , 
\end{equation}
where $\alpha$, $\lambda$, $\xi$ and $\sigma $ are constant
parameters. 
$K_{ij}$ is the extrinsic curvature, defined by
\begin{equation}
K_{ij} = \frac{1}{2N} (\dot g_{ij} - \nabla_i N_j - \nabla_j N_i) ,
\end{equation}
where the dot denotes the partial derivative with respect to $t$.
For the Ho\v{r}ava-Lifshitz gravity, the parameters are given in terms 
of the original parameters by Ho\v{r}ava~\cite{horava2009_lifshitz}
due to the detailed balance condition 
as~\cite{kiritsis_kofinas2009}
\begin{equation}
\alpha=\frac{2}{\kappa^{2}},\quad \xi=\frac{\kappa^{2}\mu^{2}}{8(1-3\lambda)}\Lambda_{W},
\quad \sigma =\frac{\kappa^{2}\mu^{2}}{8(1-3\lambda)}(-3\Lambda_{W}^{2}).
\end{equation}
Although this sign of $\sigma$ implies the negative cosmological
constant for $\lambda>1/3$, this can be flipped by making the analytical
continuation of Ho\v{r}ava's original parameters~\cite{lmp2009}.
We can also include the vacuum energy contribution from the 
matter sector and make the total cosmological constant positive.  
Moreover, the above IR limit is recovered not only by the Ho\v{r}ava-Lifshitz gravity
but also by much wider class of nonrelativistic quantum gravity
theories.
In the following we mainly choose $\lambda=1$ to recover the apparent
form of general relativity and hence the apparent Lorentz invariance.
Then, we can directly compare this action to that of general relativity. 
Then, the speed of light $c$, Newton's constant $G$ and the cosmological
constant $\Lambda$ are related to the parameters as follows:
\begin{equation}
\alpha= \frac{1}{16\pi G c}, \quad  \xi=c^{2}\alpha , \quad \sigma =-2\Lambda c^{2} \alpha.
\label{eq:parameters}
\end{equation}
We can also consider the matter action but here we focus on the vacuum
case except for the possible contribution to the cosmological constant.

To obtain the field equations we take the variation of the action 
with respect to the lapse $N$, the shift $N^{i}$ and 
the spatial metric $g_{ij}$.
To do this we should recall that Ho\v{r}ava~\cite{horava2009_membranes,horava2009_lifshitz} 
imposes the projectability condition on the lapse function $N$.
This demands that 
the lapse function $N$ be a function only of $t$, i.e., $N=N(t)$,
while the shift $N^{i}$ and the spatial metric $g_{ij}$ are allowed 
to be functions of both $t$ and $x^{i}$.
The projectability condition is favourable from a quantisation
point of view because the Hamiltonian and momentum constraints then constitute
a Lie algebra with respect to the Poisson brackets unlike in general 
relativity~\cite{horava2009_membranes,horava2009_lifshitz}.

The variation with respect to $N$ implies
\begin{eqnarray}%
{\cal H}_{0}\equiv 
\int d^3{\bf x} \sqrt{g}\left[-\alpha (K_{ij}K^{ij}\!-\!\lambda K^2)\!+\!\xi
R\!+\!\sigma\! 
\right]=0.
\label{hamiltonian}
\end{eqnarray}%
Thus, due to the projectability condition,
the Hamiltonian constraint is given in terms of the spatial integral.
On the other hand, the variation with respect to 
$N_{i}$ implies the momentum constraint
\begin{equation}
H^{i}\equiv 2\alpha(\nabla_j K^{ji}-\nabla^i K)
=0,\label{eom2}
\end{equation}
which is a local equation.

The equations obtained through the variation with respect to $ g_{ij}$, which 
are the evolution equations, are the following:
\begin{eqnarray}
&& \alpha \left[\frac{1}{2}K^{lm}K_{lm}g^{ij}-2K^{im}K^{j}_{m}
-\frac{1}{N\sqrt{g}}(\sqrt{g}K^{ij})^{\cdot} 
-\nabla_{p}\left(K^{ip}v^{j}\right)-\nabla_{p}\left(K^{pj}v^{i}\right)
+\nabla_{p}\left(K^{ij}v^{p}\right)\right]\nonumber \\
&& -\alpha\left[
\frac{1}{2}K^{2}g^{ij}-2KK^{ij}-\frac{1}{N\sqrt{g}}
(\sqrt{g}Kg^{ij})^{\cdot} -\nabla_{p}\left(Kg^{ip}v^{j}\right)
-\nabla_{p}\left(Kg^{jp}v^{i}\right)
+\nabla_{p}\left(Kv^{p}g^{ij}\right)\right]\nonumber \\
&& +\xi 
\left[\frac{1}{2}g^{ij}R-R^{ij} \right]
+\sigma \frac{1}{2}g^{ij}
=0,
\label{eq:full_evolution_equation_projectable}
\end{eqnarray}
where $v^{i}\equiv N^{i}/N$.

We here assume that the spacetime is static, spherically symmetric and vacuum.
The line element on the $t=$const spacelike hypersurface 
in the areal coordinates is given by 
\begin{equation}
ds^{2}=e^{2\omega(r)}dr^{2}
+r^{2}(d\theta^{2}+\sin^{2}\theta d\phi^{2}).
\end{equation}
The shift vector $N^{i}$ is required to satisfy 
$N^{r}=N^{r}(r)$, $N^{\theta}=N^{\phi}=0$ in these coordinates.

Then, the momentum constraint yields
\begin{eqnarray}
H_{r}=2\alpha (\nabla_{j}K^{j}_{r}- \nabla_{r}K) =-4\alpha \frac{1}{r}\omega'v^{r}=0,
\end{eqnarray}
where the prime denotes the derivative with respect to $r$.

The $(i,j)=(r,r)$ and $(\theta,\theta)$ components 
of the evolution equations respectively yield
\begin{eqnarray}
&& \alpha\frac{v^{r}}{r}\left(2v^{r\prime}+\frac{v^{r}}{r}+4\omega'v^{r}\right) 
+\xi \frac{1-e^{-2\omega}}{r^{2}}+\frac{1}{2}\sigma=0, 
\label{eq:rr}\\
&& \alpha \left[\left(v^{r\prime}+\omega'v^{r}+\frac{v^{r}}{r}\right)' v^{r}
+(\omega' v^{r}+v^{r\prime})
\left(v^{r\prime}+\omega'v^{r}+\frac{v^{r}}{r}\right)
 +\left(\frac{v^{r}}{r}\right)^{2}\right] \nonumber \\
&& ~~+\xi \frac{\omega'}{r}e^{-2\omega}+\frac{1}{2} \sigma=0
\label{eq:thetatheta}
\end{eqnarray}
The above two equations give all independent components of 
the evolution equations.
\section{III. ~STATIC SPHERICALLY SYMMETRIC VACUUM SOLUTIONS}

We can easily solve the momentum constraint. The result is $\omega'=0$
or $v^{r}=0$. Hereafter, we solve the equations for each case separately.

In the first case, $\omega=\omega_{0}$, where $\omega_{0}$ is constant.
Then, Eq.~(\ref{eq:rr}) yields
\begin{equation}
\alpha \left[\frac{v^{r}}{r}\left(2 v^{r\prime}+\frac{v^{r}}{r}
\right)\right]+\xi\frac{1-e^{-2\omega_{0}}}{r^{2}}+\frac{1}{2}\sigma=0.
\end{equation}
This can be easily integrated. Thus we obtain the following solution:
\begin{equation}
v^{r2}=-\frac{\xi}{\alpha}(1-e^{-2\omega_{0}})-\frac{1}{6}
\frac{\sigma}{\alpha}r^{2}+\frac{C}{r},\quad \omega=\omega_{0}
\label{eq:solution1}
\end{equation}
where $C$ is a constant of integration. We can easily check the
above solution satisfies both Eqs.~(\ref{eq:rr}) and (\ref{eq:thetatheta}).
For this solution, we have 
\begin{equation}
R^{r}_{r}=0, \quad R^{\theta}_{\theta}=R^{\phi}_{\phi}=\frac{1-e^{-2\omega_{0}}}{r^{2}},
\end{equation}
and hence we can see that the IR limit is justified for $r\to\infty $.

Next we will check the global Hamiltonian constraint. 
For spherically symmetric case, the global Hamiltonian 
in the IR limit for vacuum is given by
\begin{equation}
{\cal H}_{0}=
4\pi \int dr r^{2}e^{\omega}
\left[-\alpha (K_{ij}K^{ij}-K^{2})+\!\xi
R\!+\!\sigma\! \right]=0.
\label{eq:spherical_hamiltonian}
\end{equation}
For the static case, we have
\begin{equation}
K_{ij}K^{ij}-K^{2}=-\frac{2}{r^{2}}(r v^{r2})'.
\end{equation}
When we substitute the solution into the integrand,
we can see it identically vanishes. 
Thus, Eq.(\ref{eq:solution1}) gives a
solution to the Einstein-Hilbert action theory with the 
projectability condition. 

It is interesting to see what this solution describes.
For any static metric which is given by the following form 
\begin{equation}
ds^{2}=-N^{2}c^{2}dt^{2}+e^{2\omega(r)}(dr+N^{r}(r)dt)^{2}+
r^{2}(d\theta^{2}+\sin^{2}\theta d\phi^{2}),
\end{equation}
where $N$ is constant,
we can always transform this into the diagonal form given by
\begin{eqnarray}
&& ds^{2}=-\left(1-e^{2\omega}\frac{v^{r2}}{c^{2}}\right)c^{2}dT^{2}
+\left(1-e^{2\omega}\frac{v^{r2}}{c^{2}}\right)^{-1}e^{2\omega}
dr^{2}+r^{2}(d\theta^{2}+\sin^{2}\theta d\phi^{2}),
\label{eq:diagonal_metric}
\end{eqnarray}
through the coordinate transformation
\begin{equation}
N dt=dT+\frac{e^{2\omega}v^{r}}{c^{2}-e^{2\omega}v^{r2}}dr.
\end{equation}

We substitute Eq.~(\ref{eq:solution1}) into
Eq.~(\ref{eq:diagonal_metric})
and recover $G$, $c$ and $\Lambda$ using Eq.~(\ref{eq:parameters}).
Rescaling $T$ as $e^{\omega_{0}}T\to T$ and 
putting  $C= 2GM$,
we can find the Kottler solution in the standard form
\begin{eqnarray}
&& ds^{2}=-\left[1-\frac{1}{3}\Lambda r^{2}-\frac{2GM}{c^{2}r}\right]c^{2}dT^{2}
+\left[1-\frac{1}{3}\Lambda r^{2}-\frac{2GM}{c^{2}r}\right]^{-1}dr^{2} 
\nonumber \\
&&~~
+r^{2}(d\theta^{2}+\sin^{2}\theta d\phi^{2}).
\end{eqnarray}
Thus, in this case, the Kottler solution is the only solution
and there is no degree of freedom to choose a functional form in 
the distribution of dark matter. 

In the second case, $N=$const, $N^{r}=0$ and $\omega=\omega(r)$ 
must satisfy the evolution equations.
This is called an ultra-static metric in the 
literature~\cite{Gibbons_etal_2009}. 
Eq.~(\ref{eq:rr}) then yields
\begin{equation}
\xi \frac{1}{r^{2}}(1-e^{-2\omega})+\frac{1}{2}\sigma=0.
\end{equation}
Therefore, we find
\begin{equation}
e^{2\omega}=\left(1+\frac{\sigma}{2\xi}r^{2}\right)^{-1}
=\left(1-\Lambda r^{2}\right)^{-1}.
\end{equation}
It is easily found that this satisfies both Eqs.~(\ref{eq:rr}) and
(\ref{eq:thetatheta}).
Thus, rescaling $t$, we obtain the following metric
\begin{equation}
ds^{2}=-dt^{2}+\frac{1}{1-\Lambda r^{2}}dr^{2}+r^{2}
(d\theta^{2}+\sin^{2}\theta d\phi^{2}).
\end{equation}
For $\Lambda>0$, $0<r<1/\sqrt{\Lambda}$ is allowed and the 
spatial geometry is given by the three-dimensional sphere of radius $1/\sqrt{\Lambda}$.
For $\Lambda<0$, the spatial geometry is given by the three-dimensional 
hyperboloid.
In fact, the above argument would not depend on the symmetry assumption.
For the general ultra-static vacuum metric, 
Eq.~(\ref{eq:full_evolution_equation_projectable}) yields 
\begin{equation}
R^{ij}=2\Lambda g^{ij}, 
\end{equation}
For the three dimensional space, this directly means that 
the space is flat, spherical or hyperbolic depending on the sign of $\sigma$. 
The above equation also implies that the IR limit is justified if $\Lambda$ is 
sufficiently small. 

However, in this case, the global Hamiltonian constraint plays an important role.
The integrand of Eq.~(\ref{eq:spherical_hamiltonian}) is calculated to give
\begin{equation}
r^{2}e^{\omega}\left[-\alpha (K_{ij}K^{ij}-K^{2})+\!\xi
R\!+\!\sigma\! \right]=-2\sigma r^{2} \left(1-\Lambda r^{2}\right)^{-1/2}.
\end{equation}
The integration yields
\begin{equation}
{\cal H}_{0}=\left\{
\begin{array}{ll}
\displaystyle\frac{c}{2G}\left[\frac{\arcsin(\sqrt{\Lambda}r )}{\sqrt{\Lambda}}-r\sqrt{1-\Lambda r^{2}}\right]^{r_{\rm max}}_{r_{\rm min}} 
& \mbox{for}~ \Lambda>0 \\
0 & \mbox{for}~ \Lambda=0 \\
\displaystyle-\frac{c}{2G}\left[r\sqrt{1-\Lambda r^{2}}-\frac{\mbox{arcsinh}{\sqrt{-\Lambda}r}}
{\sqrt{-\Lambda}}\right]
^{r_{\rm max}}_{r_{\rm min}} 
& \mbox{for}~ \Lambda<0,
\end{array}
\right.
\end{equation}
where the region of the spacetime described by this metric is 
$r_{\rm min}<r<r_{\rm max}$.
Therefore, the global Hamiltonian constraint cannot be satisfied with 
these metrics alone except for $\sigma=0$, for which the spacetime 
is Minkowski.
On the other hand, since the Hamiltonian constraint is global, it is 
still possible that the ultra-static metric with $\Lambda\ne 0$ 
obtained here can describe a vacuum region which is part of the whole 
spacetime, where the nonvanishing contribution from the region
described by the present metric is compensated by the 
contribution from the region which is not vacuum and/or not static.

\section{IV. ~DISCUSSION AND CONCLUSION}

We have investigated that static spherically symmetric vacuum solutions to 
the IR limit of nonrelativistic quantum gravity with the 
projectability condition. We have shown that such solutions are 
uniquely given by the Kottler solutions if the spacetime is entirely
empty. If the spacetime is instead locally vacuum, we can also 
have the possibilities that the spacetime is locally given by 
the Minkowski or ultra-static solutions with sphere or hyperboloid 
according to the sign of the cosmological constant.
In the latter case, the nonvanishing contribution to the global 
Hamiltonian constraint must be compensated by the contribution 
from nonempty or nonstatic regions of the spacetime.

Mukohyama~\cite{mukohyama2009} showed that 
the IR limit of nonrelativistic quantum gravity
can be regarded as general relativity plus a dust fluid,
which emerges not from the matter sector but as a constant of 
integration due to the nonlocal Hamiltonian constraint and 
that the four-velocity of this dust is normal to the constant 
$t$ spacelike hypersurface compatible with 
the projectability condition.
We call this dust `dark dust' to make the discussion clear. 
It is true that the dark dust plays the same role as dark matter 
in the homogeneous universe. On the other hand, we have seen that 
the static spherically symmetric spacetime is uniquely described by the Kottler
solution if the spacetime is entirely empty, in spite of the 
proper implementation of the projectability condition.
This means that the theory in spherical symmetry 
would not have the freedom of choosing the mass function 
of the dark dust to exhibit the flat rotation curve observed in galaxies.
Moreover, since the four-velocity of the dark dust is 
given by the unit normal of 
constant $t$ hypersurfaces, i.e., 
$u_{a}=c n_{a}=-c N(t)\nabla_{a}t$,
the dark dust cannot rotate and hence cannot 
have the stationary configuration with rotating.
It would be also impossible that the dark dust obtains 
some kind of velocity dispersion which makes the
static configuration possible.
As a result, it would be very unlikely for the dark dust 
in this category of theories can explain galactic dark matter
responsible for flat rotation curves.  

The uniqueness of the static spherically symmetric
vacuum solution, which holds for locally vacuum regions 
in general relativity, would not hold in the present theory. 
The uniqueness in the present case is the following:
1) If the spacetime is static, spherically symmetric and 
vacuum everywhere, the spacetime 
is given by the Kottler solution.
2) If the the spacetime is static, spherically symmetric and 
vacuum within a spherical shell of finite thickness, the spacetime
there is given by the Kottler solution or 
the static three-dimensional sphere or the static three-dimensional 
hyperboloid, depending upon the sign of the cosmological constant.
The projectability condition clearly plays an important role
in this conclusion.

It has been believed that the IR limit of Ho\v{r}ava-Lifshitz 
gravity or other nonrelativistic quantum gravity theories
recovers general relativity. On the contrary, the existence of 
the dark dust clearly questions this belief. Moreover, 
as we have seen, the uniqueness of the vacuum static spherically symmetric
spacetime is quite different from that in general relativity.
As a result, we conclude that Ho\v{r}ava-Lifshitz gravity as well as
nonrelativistic quantum gravity with the projectability condition 
would not recover general relativity in their IR limit. 
It should be however noted that this fact does not immediately
mean that the theory is not viable. It is very important to study 
whether these features which are different from general relativity in the 
IR limit would leave any difference which is experimentally testable.
It can be conjectured that all solutions of general relativity 
are also solutions of the IR limit of nonrelativistic quantum gravity 
theories with the projectability condition but not vice versa.

It should be noted that there generally 
appears a spin-0 scalar mode of gravitational 
waves in the Minkowski background in 
this category of theories. This mode is potentially dangerous 
from a phenomenological point of view~\cite{bogdanos_saridakis_2009,wang_maartens_2009,svw2009_letter,svw2009_full}.

After this paper was submitted, the authors realised that Tang and 
Cheng~\cite{tang_chen_2009} reported apparently similar results. 
The present paper focused the uniqueness of the solutions and 
is complementary to their work.\\
\\
{\bf Acknowledgements:}

The authors would like to thank J. Soda, S. Yahikozawa, 
K. Nakao, T. Kuroki, S. Mukohyama  and M. Saijo for 
valuable comments and discussion. 
TH was supported by the Grant-in-Aid for Scientific
Research Fund of the Ministry of Education, Culture, Sports, Science
and Technology, Japan [Young Scientists (B) 21740190].

\end{document}